\documentstyle[11pt,newpasp,twoside,epsf]{article}
\makeindex
\begin{document}
\title{A $\gamma$ Doradus star campaign}
 \author{ Laurent Eyer}

\affil{ Princeton University Observatory,
 Peyton Hall, Princeton, NJ 08544, USA }
\author{ Conny Aerts, Mieke Van Loon, Filip Bouckaert }

\affil{ Instituut voor Sterrenkunde,
 Katholieke Universiteit Leuven,
 Celestijnenlaan 200 B, B-3001 Leuven, Belgi\"e }
\author{Jan Cuypers}\affil{Koninklijke Sterrenwacht van Belgi\"e, Ringlaan~3,
 B-1080 Brussels, Belgium}
\begin{abstract}

We report on the results from a large photometric campaign on thirty five
$\gamma\,$Dor star candidates undertaken at the Institute of Astronomy of
the University of Leuven in the framework of the Flanders -- South-Africa
grant. An overview of the data, as well as the results of the analysis of
the obtained time series are presented, the main conclusion being that
nine stars are thought to be multiperiodic $\gamma\,$Dor stars and eight
monoperiodic.
We also performed a photometric mode identification for two stars of
the sample by comparing the amplitude ratios in the different passbands
of the Geneva photometric system. Both stars seem to pulsate in
non-radial modes of degree $\ell=1$.

\end{abstract}

\section{Introduction}

\index{$\gamma\,$Dor stars}
This story starts with Cousins (see the review by Kilkenny -- these
proceedings) \& Warren's (1963) suggestion of having found photometric
variability in the F0\,V star $\gamma\,$Dor. \index{$\gamma\,$Dor} Cousins (1966) himself
confirmed unambiguously his discovery of the variability in
$\gamma\,$Dor, the star that later would become the prototype of a new
class of pulsating stars. Much later, Cousins (1992) was able to
detect two close periods in $\gamma\,$Dor~: 0.7570\,d \&
0.7334\,d.

Balona et al.\ (1996) undertook a very detailed photometric and
spectroscopic follow-up study of $\gamma\,$Doradus and found one
additional frequency. Moreover, they refined the frequency values to
1.32098\,d$^{-1}$, 1.36354\,d$^{-1}$, 1.47447\,d$^{-1}$. For illustrative purpose, we
show in Fig.~1 the phase diagrams for the three detected
frequencies of both the photometric and spectroscopic data of Balona
et al.\ (1996). Earlier on, Balona et al.\ (1994) already proposed
$\gamma\,$Dor to be a member of ``a new class of non-radial\index{non-radial pulsation}
gravity-mode pulsators''. With their 1996 paper, Balona et al.\ left
no longer doubt about the cause of $\gamma\,$Dor's variability: the
radial-velocity variations excluded binarity, while spot models were
also excluded since: 1) they required a too large differential
rotation, 2) the periods were too stable, and 3) there was no phase
difference between the radial-velocity and light maxima. It therefore
seemed unavoidable to conclude that $\gamma\,$Dor pulsates
multiperiodically, with pulsation periods an order of magnitude longer
than those of pressure modes.

\begin{figure}
\begin{center}
\plotone{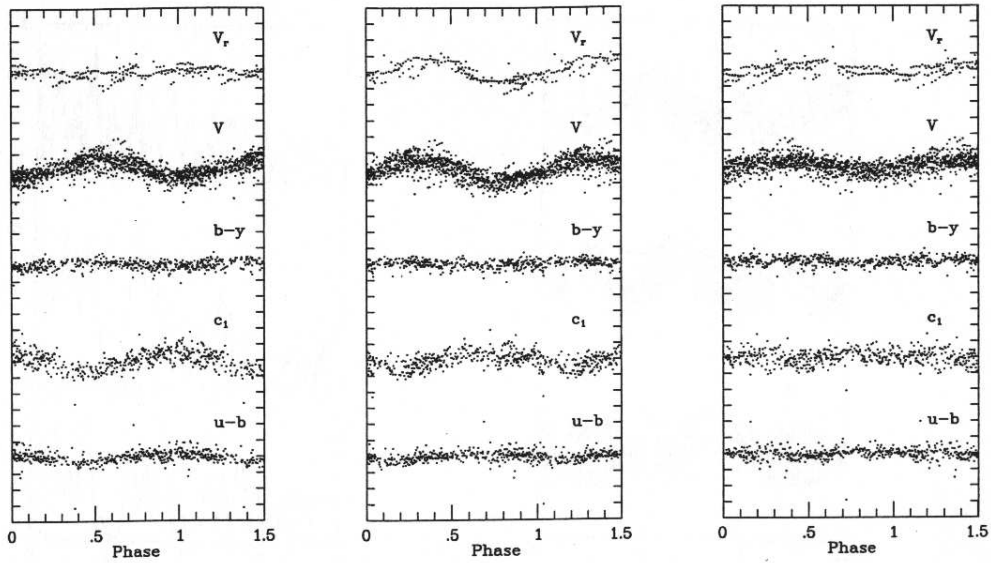}
\end{center}
\caption{\label{fig:gdor} Phase diagrams of the radial-velocity, light and
colour variations of the prototype of the class of $\gamma\,$Dor stars for the
frequencies 1.32098\,d$^{-1}$ (left), 1.36354\,d$^{-1}$ (middle), 1.47447\,d$^{-1}$ (right). Figure taken from Balona et al.\ (1996) with permission. }
\end{figure}

Meanwhile, several comparison stars for other variables turned out to
have similar variability to $\gamma\,$Dor:
\begin{itemize}
 \item HD\,164615 (Burke et al., 1977);\index{HD\,164615}
 \item HD\,55892 (Hensberge et al., 1981);\index{HD\,55892}
 \item four variables in NGC\,2516 (Antonello \& Mantegazza, 1986);\index{NGC\,2516}
 \item 9\,Aur (Krisciunas \& Guinan, 1990);\index{9\,Aur}
 \item HD\,111828 (Mantegazza et al., 1991);\index{HD\,111828}
 \item HD\,224638 \& HD\,224945 (Mantegazza \& Poretti, 1991);\index{HD\,224638}\index{HD\,224945}.
\end{itemize}

This led Krisciunas \& Handler (1995) to publish a list of 17
candidate $\gamma\,$Dor stars. Since the publication of the list by
Krisciunas \& Handler, several additional candidates were found from
ground-based photometry (Henry, 1995, 1999; Breger et al., 1997; Kaye
et al., 1999; among others). Presently, an updated list is maintained
by Handler at {\tt http://www.saao.ac.za/$\sim$gerald/gdorlist.html}.

It should be noted that the periodic variability of only very few
stars is illustrated by phase plots of the quality shown in
Fig.~\ref{fig:gdor}. For most stars the scatter is considerably larger.

The cause of the non-radial gravity modes\index{gravity modes} is still a matter of
debate. Guzik et al.\ (2000) have proposed an excitation mechanism,
but this certainly is not generally accepted (see, e.g., Wu 2002;
L\"offler 2002). It is therefore of importance to find as many members
as possible and to derive their basic physical stellar parameters,
besides their frequencies and mode identification.

One of the early ideas about the general properties of $\gamma\,$Dor
stars is that such pulsators would be young. This suggestion was based
on the fact that circumstellar matter was discovered around some
member stars, among which is $\gamma\,$Dor itself. Inspired by this
proposal, searches for $\gamma\,$Dor stars in clusters of different
ages were undertaken. Krisciunas \& Patten (1998) discovered 2
$\gamma\,$Dor candidates (out of 15 stars) in M\,34
(age\,$\simeq$\,195\,Myr), while Zerbi et al.\ (1998) came up with 8
candidates (out of 44 stars) in NGC\,2516 (age\,$\simeq$\,107\,Myr). An
enormous effort in this respect was done by Martin (2000), who checked
149 stars in 9 clusters. She found only 3 $\gamma\,$Dor stars: 2 in
the Pleiades (age\,$\simeq$\,83\,Myr) and 1 in Coma Ber
(age\,$\simeq$\,490\,Myr). The conclusion therefore must be that current
studies do not reveal a clear relationship between age and
the $\gamma\,$Dor phenomenon.

The FlanSA $\gamma\,$Dor campaign found its origin at the IAU General
Assembly in Kyoto in 1997, when CA convinced LE to come to Leuven for
a scientific visit to search in detail for new $\gamma\,$Dor stars in
the {\sc hipparcos} data base. LE's visit resulted in a chain of
initiatives with respect to the study of $\gamma\,$Dor stars, which we
highlight in this paper. First we report on the status of the FlanSA
$\gamma\,$Dor campaign in Section\,3, after having highlighted the
contribution into this field by the {\sc hipparcos} mission
(Section\,2). The analyses of the time series are described in
Section\,4, while an attempt to identify the modes in two stars is
explained in Section\,5. We end this paper with future prospects and
additional current studies of $\gamma\,$Dor stars at the Institute in
Leuven and worldwide.

\section{The Contribution of the {\sc hipparcos} mission}
We recall that the {\sc hipparcos} photometry provides a homogeneous all-sky
survey down to $V$ magnitude 7.3-9.0 (depending on the galactic
latitude and the spectral type of the observed star). The {\sc hipparcos}
satellite took 13 million photometric measurements of about 118\,000
stars over a period of 3.3 yr (52\,000 stars are from the survey and
66\,000 stars were selected objects). The {\sc hipparcos} photometry \index{{\sc hipparcos}}proved
to be very useful to reveal large numbers of new variables of many
different kinds (see ESA 1997: fields H6, H49-54 of the main catalogue
and volumes 11 and 12). In specific cases the number of known variable
stars increased dramatically, e.g. the number of Slowly Pulsating B stars increased
nearly by a factor 10 (Waelkens et al. 1998).

{\sc hipparcos} has rather good detection capabilities for determining
correctly the characteristic periods of $\gamma\,$Dor stars (0.5-3~d). The sampling does not suffer from the alias problems of Earth
observations for such stars. Indeed, the spectral windows of {\sc hipparcos}
and ground-based data show very different patterns. However, the
precision of photometric individual measurements degrades quite
rapidly for fainter stars (although more time was allocated to observe
them), impeding the unambiguous variability detection of fainter
objects.

Another difficulty with the {\sc hipparcos} data occurred specifically for
$\gamma\,$Dor stars, since their variability behaviour is not always
strictly periodic. This, in combination with the rather low density
of data points of the {\sc hipparcos} sampling, prohibited an easy search
for new candidate members of the class. For instance, when we compare
the bona-fide member list (which contained 13 stars in the version of
2000) and the {\sc hipparcos} catalogue, we find that all stars listed were
measured by {\sc hipparcos} and were classified as follows: 4 stars are in
the Variability Annex as periodic variables, one star as unsolved
variable, three stars are flagged as microvariable, four are without
classification and one flagged as constant. A remark worth making is
that the four stars in the Periodic variables of the Variability Annex
have periods at the short end of the interval 0.5-3~d.

Eyer (1998) and Eyer \& Grenon (1998) did a selection of $\gamma\,$Dor
candidates, which was made from the published Periodic variable star
Annex from the {\sc hipparcos} catalogues (volume 11 and 12, ESA 1997). This
preliminary study was based on stars which had precise parallax and
colour $B-V$. In the HR diagram, there was clearly at the cool edge of
the region occupied by the $\delta\,$Scuti stars \index{$\delta$ Scuti stars}a very clumped group
of stars with periods and amplitudes compatible with the $\gamma\,$Dor
stars and thus these were declared good candidate $\gamma\,$Dor stars.

Aerts, Eyer \& Kestens (1998), in an effort to be more general and to
settle better the case of membership, performed a multivariate
discriminant analysis by using Geneva photometric indices
in order to determine the physical properties of the stars. As a
supplementary test, all stars were checked against
multiperiodicity. The stars were drawn from the Periodic variable star
Annex. The two first studies listed the name of 29 stars identified with
{\sc hipparcos} photometry (three being redundant with the list of Handler
\& Krisciunas 1997).

Because of the sometimes irregular nature of $\gamma\,$Dor light
curves, candidates could have been rejected from the Periodic variable
Annex by the {\sc hipparcos} groups analyzing and producing the catalogues.
Handler (1999a) decided to go through the {\sc hipparcos} data once more,
but looked at the periodic variability Annex as well as at the
Unsolved catalogues. He also widened the criteria of acceptable
spectral types. His analysis permitted the addition to the list of 31
non-redundant prime candidates.

We recall here that only 6 years ago, the number of total objects was
17 (6 bona fide members and 11 candidates -- Krisciunas \& Handler 1995). As
we can see from the successive lists provided in publications and/or on
the Internet, this field has undergone intense studies. At the time of
this oral presentation there were 13 bona-fide members listed by
Handler, and the number of candidates amounted 122\footnote{Since then
the list maintained by Gerald Handler has been updated, and we have the
following progression of respectively bona fide members and candidates
for different years
1995: (6, 11);
1997: (9, 16);
2000: (13, 122);
2001: (24, 120).
}.

We assisted in the explosion of the number of candidates and therefore
undertook the initiative of settling the origin of the variability for
each candidate $\gamma\,$Dor star. This somewhat unrewarding task of
removing objects from a list of potentially very exciting stars
requires extensive new photometric campaigns. The Institute of
Astronomy in Leuven decided to undertake this effort. C. Aerts took
the initiative of applying for several weeks of observing time in
South Africa in the framework of the FlanSA project. L.\ Eyer
became the driving force of the planning and exploration of the observing time
allotted in Sutherland. At the same time the 1.2-m Swiss \index{Euler
telescope}at La Silla was used to monitor each of the candidates
spectroscopically. Furthermore the Leuven 1.2-m telescope Mercator \index{Mercator telescope}(a
twin of Euler) at La Palma has become operational (since a few months prior to this writing) 
and is currently used to monitor the northern $\gamma\,$Dor candidates
photometrically in the seven-colour Geneva system. We here report on
the results derived from the data gathered in South Africa.

\section{The observational Campaign at SAAO}
%===========================================

The time allocated for this project was 15 weeks spread over one year
(from 1999 October to 2000 September) in order to observe 35 stars on
the 0.5-m telescope located in Sutherland. The observations were done
in the $B$, $V$, $I$ bands in the sequence $I-V-B-B-V-I$ (30-s
exposures in each filter) and then sky measurements in $I-V-B$ (10~s
in each filter). The whole sequence for observing an individual star
took about 3 min and 50 s.

The photometry obtained is differential. Two stars were chosen as
comparison stars for each programme star. Their selection was done
using {\sc hipparcos} epoch photometry. Particular attention was paid
to remove stars with peculiar chemical composition. For those stars
the variation of the luminosity is a function of the wavelength. More
precisely, for the cooler Ap stars, the variations can be in
anti-phase in blue and red light and the signal is washed out in a
broad band filter like the $H\!p$ filter. Therefore the star appears
to be a constant. The physical reason of this is that the blocking
effect is more severe in the blue, while the re-emission is in the red
for the cool Ap stars \index{Ap stars}(in the case of hot Ap stars the
blocking effect is in the ultraviolet and the backwarming in the
visible). The comparison stars were generally chosen to be at a
distance less than 7$^{\circ}$ and of similar magnitude. Furthermore,
one comparison star was chosen to be slightly bluer than the programme
star and the other slightly redder.

The measurements consisted of three block sequences of $A-P-B$ where
$P$ is the programme star and $A$ and $B$ are the comparison stars.
Typically a star was measured 2-4 times per night. However the
sampling was sometimes denser with short sequences $A-P-B-P-A$\ldots
(where $P$ is the programme star and $A,B$ the two comparison stars)
in order to be able to distinguish periods of $\delta\,$Sct stars
from periods of $\gamma\,$Dor stars.

Some tests were also done to take photometry only in two bands at the
beginning or end of the night (before and after the astronomical
night) focusing on only one star.

\subsection{Weather conditions}
\begin{figure}
\plotfiddle{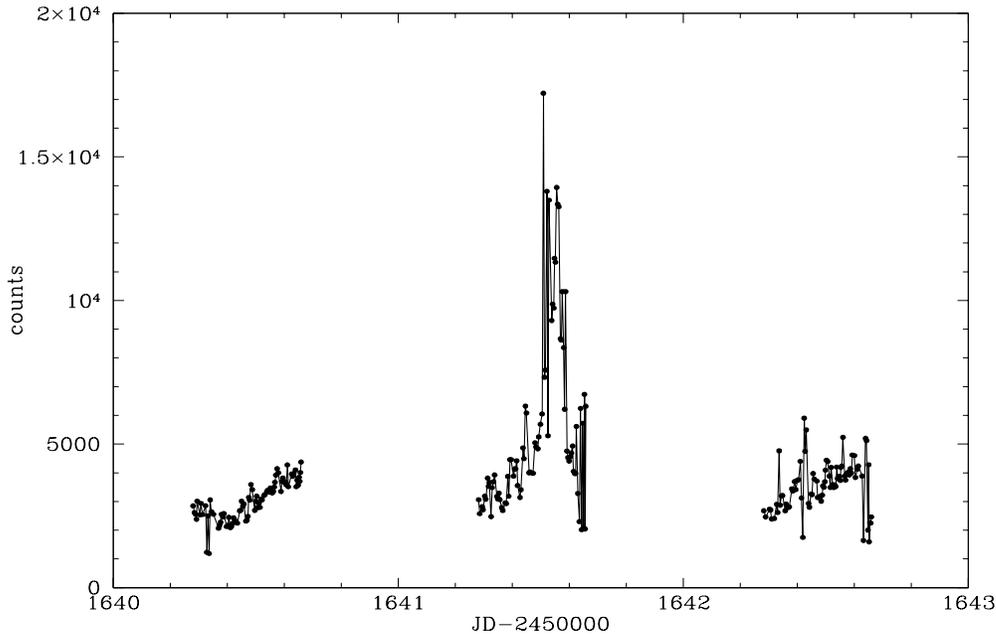}{7.8cm}{-90}{50}{45}{-200}{255}
\caption{\label{fig:aurora}
 On the night of 2000 April 6-7, a beautiful Aurora Australis
 occurred. The sky counts in the $V$ band are represented for
 three nights. The variation of the sky counts due to the aurora
 reaches nearly 2 mag. The following nights are also slightly
 perturbed.
 }
\end{figure}

%=============================
The campaigns in Sutherland suffered from relatively bad weather. 
Normally the fraction of photometric nights is about 55\%, while we had about
40--45\%. There were a few exceptional and rare events: 

\begin{itemize}

\item the tail of the bad weather causing the tragic flood in Zimbabwe
 surprisingly reached Sutherland,

 \item a beautiful Aurora occurred, which had the sad effect of rendering
 void a large part of a night. The time scale of the light
 variability caused by this aurora is short as seen in
 Fig.~\ref{fig:aurora}. We recall here that Sutherland is not
 very far South (latitude is $-32^\circ$).
\end{itemize}

%----------------------------- FIG. 2 -------------------------------------

%--------------------------------------------------------------------------

\subsection{The observers}
%=========================

We give below by affiliation the list of observers, with (in parentheses)
the number of weeks observed. From the Katholieke Universiteit Leuven
(KUL): L.~Eyer (6), P.~De~Cat (3), from the South African Astronomical
Observatory (SAAO): F.~van Wyk (2), G.~Handler (1), from the University Cape Town (UCT): M.~M\"uller (2), M.~Hempel (1). The people having no
experience with the 0.5-m received an introduction by either D.~Kurtz
or G.~Handler.

\subsection{The programme stars}
%==============================

The initial target list contained thirty-seven stars. However, at the
beginning of the observing run two stars were removed from the list
thanks to the cross-check of the list by G.~Handler (1999b). Their
$V$ magnitudes range from 5.3 to 9.5. Five stars needed to be observed
with neutral density filters. The spectral types range from A1
to F4, but most stars are classified as F0 or F2. Altogether 105
stars (programme stars and comparison stars) were observed during the
observing run. Finally we note that twenty-six of the stars are being
measured with the Echelle spectrograph {\sc coralie} \index{{\sc coralie} }attached to the Swiss
Euler telescope located at La Silla. The spectral analysis is still in
progress, so we will not refer to it here.

Thirty stars were analyzed in total in the following section. The
number of observations per star varies between 14 and 97 in the $V$
band. Seventeen stars were from the lists of Eyer (1998) and Aerts et
al.\ (1998), eleven from Handler (1999a), one from the bona fide member
list, and one from Paunzen et al.\ (1998).

\section{The analysis of the time series}

%========================================

The analysis of the time series was performed by F.~Bouckaert in the
framework of his master thesis in Leuven and was supervised by
J.~Cuypers and C.~Aerts. For those stars which are also present in
the Geneva data base, a general comparison was done with Geneva
photometry. For all stars, the variability in the South African data
was cross-checked also with the {\sc hipparcos} data and with results from
the literature, whenever possible. There is a point worth mentioning
here: in Simbad the origin of the discovery of the variability is not
reported and sometimes it is difficult to trace back the source of the
claimed variability and subsequently check more carefully the
data.

% Conny I would remove this sentence (LE): For our work, this turned out

% to be a clear disadvantage.

The period searches were done with different programs and methods,
both based on the PDM technique and on Fourier analysis. Moreover, we
used the programme developed by J.~Cuypers which fits simultaneously
several frequencies. Special emphasis was put on the search for
multiperiodicity for every star.

The stars were classified in several categories: multiperiodic
$\gamma\,$Dor stars, singly periodic $\gamma\,$Dor stars,
non-$\gamma\,$Dor stars ($\delta\,$Sct stars, eclipsing binaries)
and dubious cases. We now sum up the results for each of these
categories.

\subsection*{Multiperiodic $\gamma\,$Dor stars}

%=============================================

Our analysis resulted in the classification of 9 targets as multiperiodic $\gamma\,$Dor stars: HD\,12901 (HIP\,9807), HD\,34025 (HIP\,24215), HD\,48501 (HIP\,32144), \\ 
HD\,65526 (HIP\,39017), HD\,110606 (HIP\,62105), HD\,112685 (HIP\,63372),\\
 HD\,135828 (HIP\,74825), HD\,206481 (HIP\,107443), and HD\,209295 (HIP\,108976). \index{HD\,12901}\index{HD\,34025}\index{HD\,48501}\index{HD\,65526}\index{HD\,110606}\index{HD\,112685}\index{HD\,135828}\index{HD\,206481}\index{HD\,209295}

\begin{figure}
\plotone{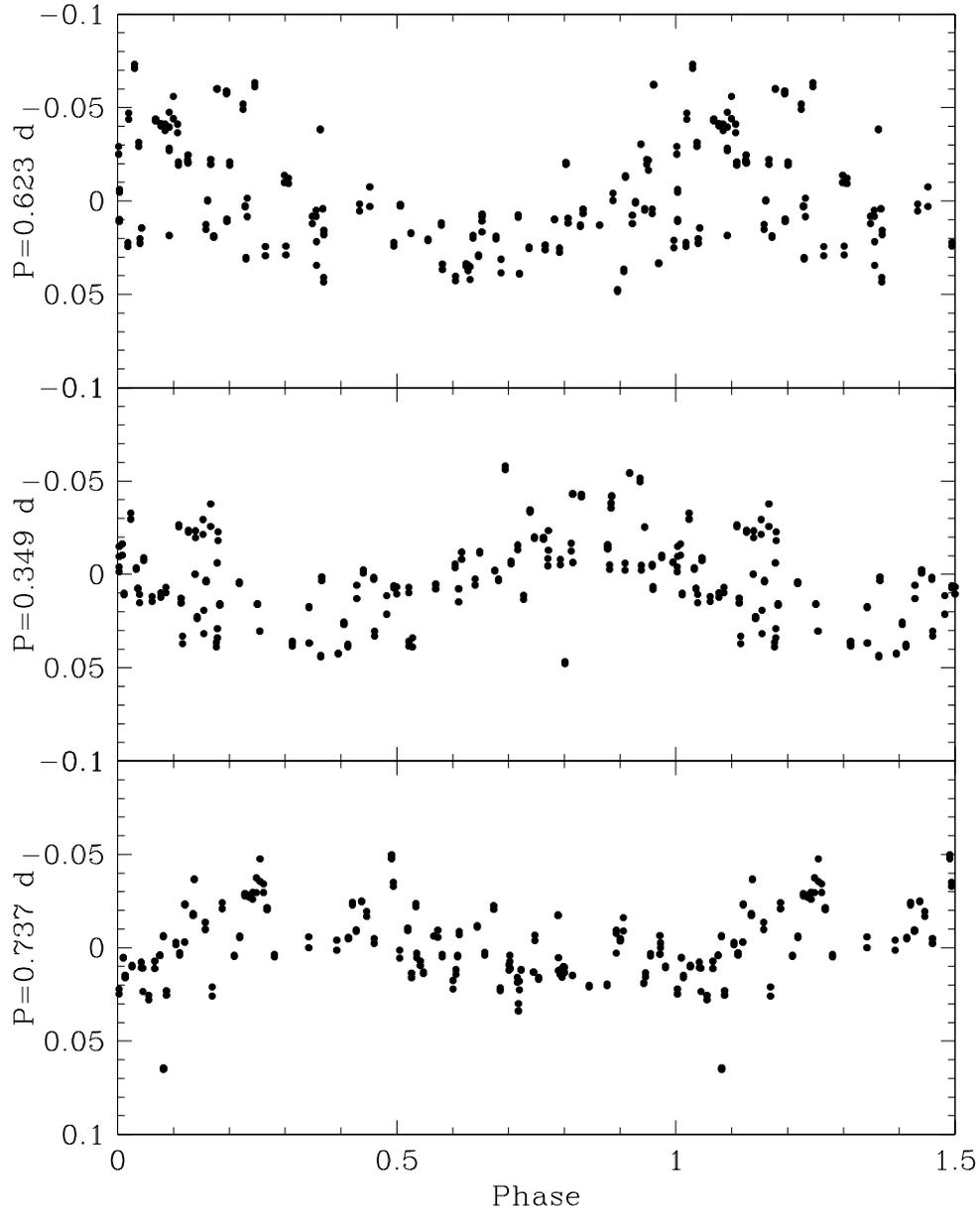}
\caption{\label{fig:multiper}
 A multiperiodic $\gamma\,$Dor star, HD\,112685 (HIP 63372).
 }
\end{figure}

As an illustration, we present in Fig.~\ref{fig:multiper} the data for
the star HD\,112685. It is clearly multiperiodic, with the
periods 0.623\,d (main period), 0.349\,d and 0.737\,d. Although we
have far fewer data points, the quality of the phase diagrams is
similar to the one of the prototype shown in Fig.~\ref{fig:gdor}. The
{\sc hipparcos} periodic variable Annex quotes a period of 0.600\,d for
this object. The phase diagrams of the eight other stars are similar
to the ones shown here for HD\,112685.

%----------------------------- FIG. 3 -------------------------------------

\subsection*{Singly periodic $\gamma\,$Dor stars}

%===============================================

There are 8 stars classified as monoperiodic $\gamma\,$Dor stars:
HD\,10167 (HIP\,7649), HD\,14940 (HIP\,11192),
HD\,26298 (HIP\,19383), HD\,40745 (HIP\,28434),
HD\,111709 (HIP\,62774), HD\,112934 (HIP\,6349),
HD\,187028 (HIP\,97590), and HD\,197451 (HIP\,102329).
\index{HD\,10167}\index{HD\,14940}\index{HD\,26298}\index{HD\,40745}\index{HD\,111709}\index{HD\,112934}\index{HD\,187028}\index{HD\,197451}

One might be tempted to conclude that about half of the $\gamma\,$Dor
stars turn out to be monoperiodic. However, we note that the stars of
small amplitude will have a tendency to be classified as monoperiodic
$\gamma\,$Dor stars for statistical reasons, particularly when
a relatively small amount of data is available. The monoperiodicity
of our targets does not necessarily imply that the fraction of
multiperiodic $\gamma\,$Dor stars is effectively higher at higher
amplitude.

\subsection*{Eclipsing binaries}

%==============================

\begin{figure}
\plotone{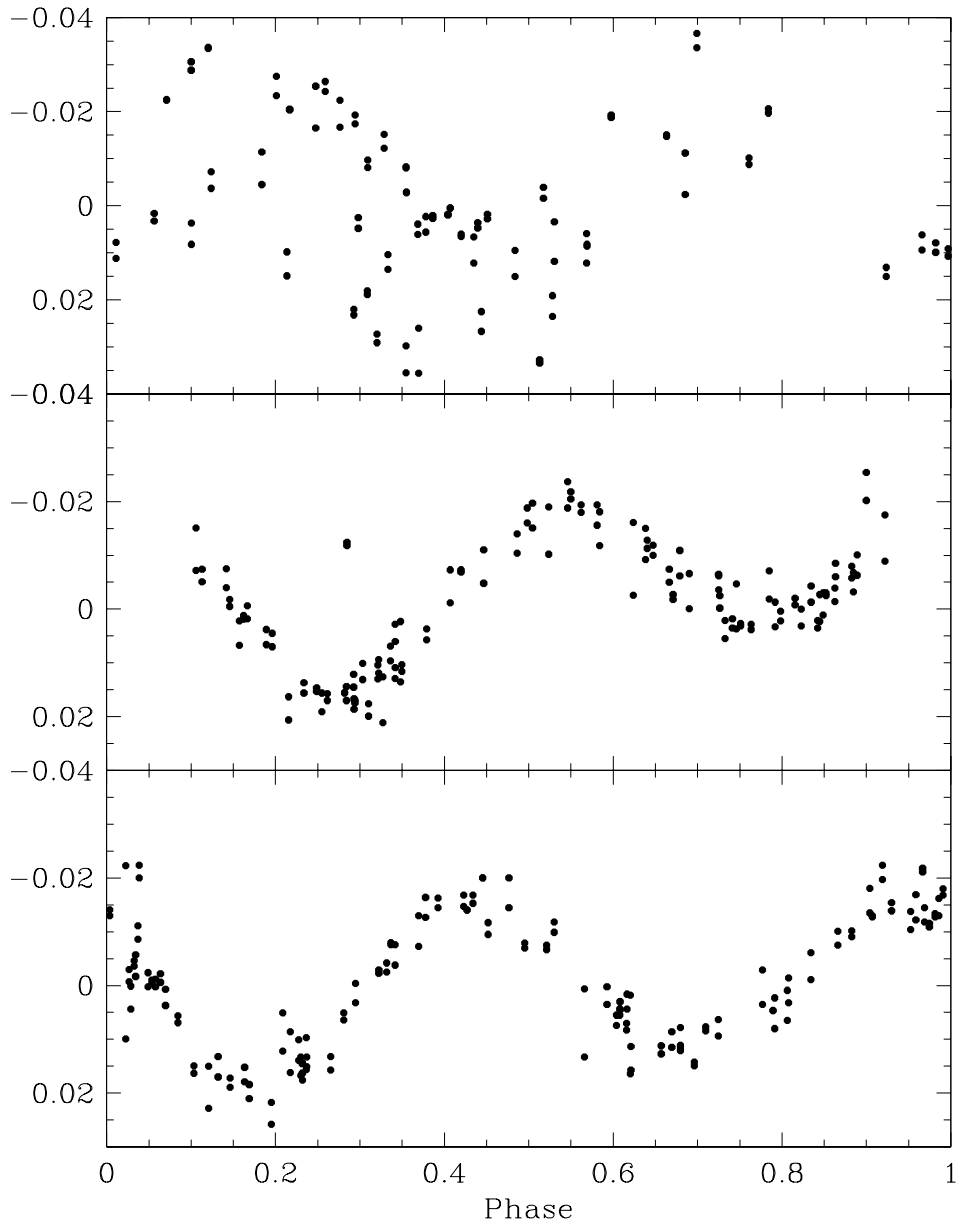}
\caption{\label{fig:eclbin}
 The phase diagrams of the three eclipsing binaries HD\,41448,
 HD\,81421 and HD\,85964 (top to bottom).
 }
\end{figure}

HD\,41448 (HIP\,28778), HD\,81421 (HIP\,46223), and\index{HD\,41448}\index{HD\,81421}\index{HD\,85964 } 
HD\,85964 (HIP\,48580) are thought to be eclipsing
binaries. This conclusion was drawn because of the difference in depth
of successive minima in the light curve, which is typical for EB type
eclipsing binaries. The phase plots of the three stars are shown in
the Fig.~\ref{fig:eclbin}.

%HD\,41448-HIP\,28778 {\sc coralie} very wide line

%HD\,81421-HIP\,46223 {\sc coralie}

%HD\,85964-HIP\,48580 {\sc coralie}

Handler (2001) would classify the star HD\,41448 as a
$\gamma\,$Dor star, though. Its light curve does indeed also support
that interpretation since it shows a clear sign of
multiperiodicity. It remains to be confirmed if the unequal minima are
true, or merely a consequence of the multiperiodicity. It may very
well be that this star is a binary with a $\gamma\,$Dor star as a
component. The spectra currently taken by {\sc coralie} will probably resolve the ambiguity.

%----------------------------- FIG. 4 -------------------------------------

%--------------------------------------------------------------------------

\subsection*{$\delta\,$Sct stars}

%=================================

\begin{figure}[t]
\plotone{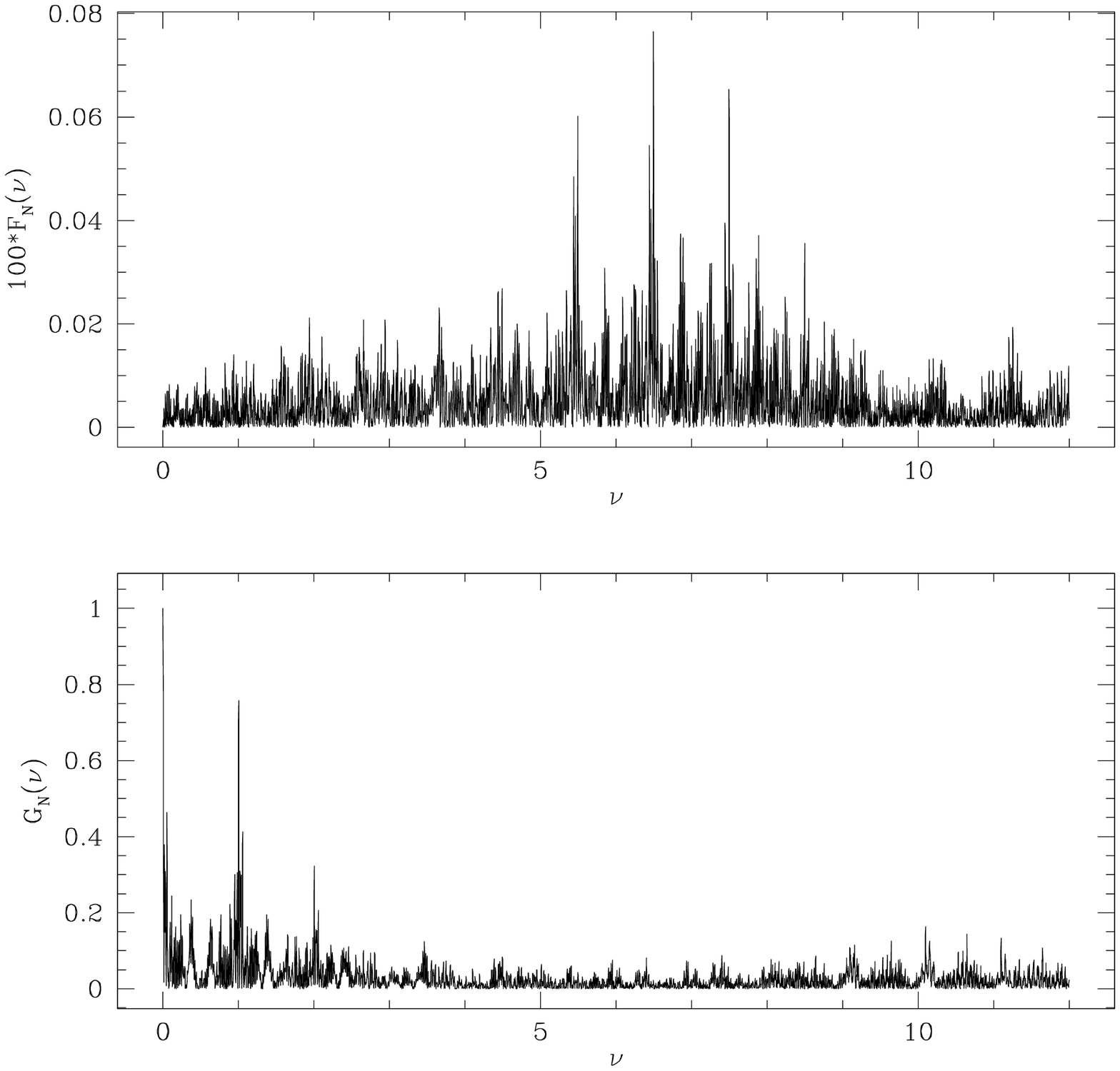}
\caption{\label{fig:deltascuti}
  Power spectrum (up) and spectral window (down) for the star HD\,125081.
  Frequencies are expressed in cycles per day.
 }
\end{figure}

There are two objects classified as $\delta\,$Sct
stars:\index{HD\,125081}\index{HD\,181998} HD\,125081 (HIP\,69848) and
HD\,181998 (HIP\,95358). The spectrum for the former star is shown in
Fig.~\ref{fig:deltascuti}. The main frequency is 6.4941\,d$^{-1}$. For
the latter object our analysis is in contradiction with the conclusion
of Handler (1999b). We find evidence for a frequency of
6.742\,d$^{-1}$, while Handler lists a typical $\gamma\,$Dor frequency
of 0.7496\,d$^{-1}$. Very probably both frequencies are aliases of
each other.

%----------------------------- FIG. 5 -------------------------------------

%--------------------------------------------------------------------------

\subsection*{Dubious cases and additional comments}

%==================================================

For three additional stars our study did not lead to firm results,
mostly because of a lack of data. This concerns HD\,171813
(HIP\,91412), HD\,197187 (HIP\,102217), and  HD\,216910
(HIP\,113402). For the latter two stars our analysis does not confirm
the previously published periods found in {\sc hipparcos} (those two
stars also have few
measurements).\index{HD\,171813}\index{HD\,197187}\index{HD\,216910}

There is one star which shows longer term behaviour in our
data:\index{HD\,27377} HD\,27377 (HIP\,20036) for which we find a
frequency of 0.100\,d$^{-1}$. This result is in conflict with Handler
(1999a), who reports the two frequencies 0.3511\,d$^{-1}$ and
0.548\,d$^{-1}$ from the {\sc hipparcos} data. Additional data are
needed to understand the variability pattern for this object.  For 4
additional studied stars no conclusions were reached.

All the results presented here will be reviewed and published with the
addition of {\sc coralie} spectra which are currently being gathered and
analysed. Special care will be taken for the claim of eclipsing
binaries and for the distinction between EW eclipsing binaries and
monoperiodic $\gamma\,$Dor stars.

%%%%%%%%%%%%%%%%%%%%%%%%%%%%%%%%%%%%%%%%%%%%%%%%%%%%%%%%

%%%%%%%%%%%%%%%%%%%%%%%%%%%%%%%%%%%%%%%%%%%%%%%%%%%%%%%%

\section{Photometric Mode identification}
For two of the target stars, we also have extensive Geneva data at our
disposal. As this leads to amplitudes at seven different wavelengths,
we decided to try to identify the pulsation modes in these two
$\gamma\,$Dor stars. This work was performed by M.~Van Loon in the
framework of her Master's thesis and was supervised by C.~Aerts.

The goal of the work is to identify the non-radial pulsation modes
thanks to the different behaviour of the pulsation as observed through
different filters. This photometric method is able to discriminate
between different (low) degrees $\ell$ but cannot derive the azimuthal
number $m$. For a review of the method of photometric mode
identification we refer to Garrido (2000). Although the method has
severe limitations because of the inherent different results
introduced by using different atmosphere models (see the discussion by
Garrido), we applied it to the two stars to get a crude idea about the
pulsational degree. As there exists currently no real overall accepted
excitation mechanism, any estimate of the observed degree is welcome
for as many stars as possible.

Only two stars were considered, HD\,12901 and HD\,48501, since these
are the only ones for which we have data in many filters. They were
measured in the seven filters of Geneva photometry ($U$, $B$, $B1$,
$B2$, $V$, $V1$, $G$). These two stars were already discussed in the
article by Eyer \& Aerts (2000), who classified them as new
$\gamma\,$Dor stars. Bouckaert found three frequencies for each of the
two stars: 1.21564\,d$^{-1}$, 1.39595\,d$^{-1}$, and 2.18374\,d$^{-1}$
for HD\,12901 and 1.29054\,d$^{-1}$, 1.09408\,d$^{-1}$, and
1.19924\,d$^{-1}$ for HD\,48501.

The physical parameters of both stars were determined from the
calibration by K\"unzli et al. (1997), making use of the stellar
models published by Schaller et al. (1992) and Schaerer et
al. (1993). They are displayed in Table~\ref{EyerTab1}. We remark that
the stars are nearly twins, except for their metallicity. Indeed their
temperature, gravity, mass, and radius are equal within the
uncertainty induced by the method of determination. The time series
analysis reveals also that the detected frequencies are rather
similar, as well as their amplitude ratios. One expects therefore,
their pulsation modes to be of similar nature as well.

\begin{table}[t]
\begin{center}
\caption{\label{EyerTab1} Physical properties of two $\gamma\,$Dor stars\vspace{3mm}}
\begin{tabular}{ccccccc}
HD & HIP & $T_{\mbox{\tiny eff}}$ $[K]$ & $\log(g)$ & [M/H] & $M$ $[M_\odot]$ & $R$ $[R_\odot]$ \\   \hline
12901& 9807& 7000 & 4.5 & -0.4 & 1.5 & 1.2 \\
48501&32144& 7000 & 4.5 & -0.1 & 1.5 & 1.1 \\ \hline
\end{tabular}\end{center}
\end{table}\index{HD\,12901}\index{HD\,48501}

The method of Heynderickx et al.\ (1994) was used for the mode
identification. With this method, the amplitude ratios with respect
to the $U$ filter are concerned. The choice of the $U$ filter is
arbitrary, but was chosen in this method because it was designed
originally for applications to $\beta\,$Cep stars, which have by far
the largest amplitude in the bluest filter. The main hypotheses and
ingredients of the method are:

\begin{itemize}
 \item the unknown non-adiabatic effects are taken into account by a
 free parameter, which is determined such that the difference
 between observed and theoretical amplitudes is minimal (same
 as in the method by Garrido),
 \item rotational effects are completely neglected,
 \item the phase difference between the light curves of different bands
 is zero (unlike the method by Garrido). 
\end{itemize}

We prefer to make use of this method, as it uses the variability in
seven filters (while Garrido's method only makes use of one colour and
one phase difference). As it turned out, for both stars, we observe no
phase difference at all in the seven filters. This is quite a
surprise, since one expects to have severe phase differences due to
non-adiabaticity. However, the data are of very high quality and leave
no doubt that the phase differences are zero. In that respect, it is
better to use as many amplitude ratios as possible to derive the degree
of the mode.
The work of M.~Van Loon consisted of several steps:

\begin{enumerate}
 \item she translated the code of Heynderickx, which was written in Pascal,
 to C$++$, 
 \item she included improved numerical interpolation tools to calculate the
 derivatives of the functions (such as the flux and the limb darkening
 function) with respect to the effective temperature and the gravity,
 \item she cross-checked if her results were compatible with those by
 Heynderickx for a large number of $\beta\,$Cep stars, 
 \item she added to the code the grid of stellar atmosphere models for the
 temperature range of the $\gamma\,$Dor stars.
\end{enumerate}

For all three frequencies in both stars, the study favours very
strongly the solution $\ell=1$. All other $\ell$-values lead to
amplitude ratios very different from the observed ones. These results
remain to be confirmed, of course, preferably by spectroscopic
studies, but they do help to discriminate between future excitation
models for the $\gamma\,$Dor stars. We remark that the models by Guzik
et al.\ (2000) imply the excitation of mainly $\ell =1$ modes, which
is thus compatible with our observational results.

\section{Future prospects}
There are a number of future projects in the pipeline, both
observational and theoretical. We mention here that a large
observational campaign is being organized and performed by P.\ Mathias
\& collaborators. It consists of photometric and spectroscopic
observations in the Northern Hemisphere. This, together with our
spectroscopic study with Euler from La Silla for the Southern stars
and our photometric study with Mercator from La Palma should allow us
to constrain the observational behaviour of the $\gamma\,$Dor stars
with high precision during the forthcoming years. Moreover, the
observational results will allow a critical confrontation with the
theoretical works that are currently being worked out (L\"offler 2002;
Wu, 2002 and maybe others).

\section{Acknowledgements}

We are thankful for the warm and generous hospitality of the researchers
and staff in Cape Town and Sutherland.

\begin{question}{Breger}
Could you comment on the high $\log g$ value ($\log g=4.49$) at
$T_{eff} = 7000$~K which you derived from Geneva photometry? It
appears very high and the star should not be below the ZAMS.
\end{question}

\begin{answer}{Eyer}
In the light of {\sc hipparcos} data, Pierre North checked the $\log g$
values given by the photometric calibration by K\"unzli et al.\ (1997)
for a sample of $\delta\,$Sct stars. Indeed, the estimates of $\log g $ seem affected by a systematic trend and to be too large by about
$\log g = 0.2$. The source of this bias is neither understood nor
identified but is under investigation.
\end{answer}

\begin{question}{Aerts}
I also point out that, in general, the determination of the effective
temperature is much more accurate than the one of the gravity. This is
true for every photometric system. For B stars, the typical
uncertainty of $\log g$ is 0.2 -- 0.5, even for bright stars.
\end{question}

\begin{question}{Balona}
1) How can you tell if a singly-periodic star may not be a close
 binary rather than a $\gamma\,$Dor star?\\
2) You mentioned that two $\gamma\,$Dor stars do not show a variation
 of phase with waveband. This is rather strange since one would
 expect such a variation owing to non-adiabatic effects.
 Can you comment?
\end{question}

\begin{answer}{Eyer}
1) If the light curve had a sinusoidal shape (with minima of equal
depth) the star was accepted as a $\gamma\,$Dor star variable.
% I do not agree I checked in {\sc hipparcos} (LE): A contact binary with components

% of similar mass will usually not have such a long period.

We also looked at the colour variations, but for unambiguous answers we will
have to wait for the spectroscopy results.\\
2) Yes, I agree, that is what we expected. But that is NOT what the stars are
 telling us. We clearly find that there is no phase difference for the two 
 stars, of which we have numerous high-quality Geneva photometry.
\end{answer}

 \begin{question}{Handler}
With the increasing number of $\gamma\,$Dor candidates (17 in 1995,
$>$ 140 in 1999) and the increasing interest in these stars it has
become very difficult to keep the $\gamma\,$Dor database up to
date. I'd therefore like to encourage everybody to communicate their
new discoveries to me for immediate inclusion in the $\gamma\,$Dor
Master list. Simply send an e-mail to gerald@saao.ac.za. Thank you for
your help to keep the database up to date.
\end{question}

\newpage
 \input{psfig.sty}

\thispagestyle{empty}
\begin{figure}[b]
%\centerline{\psfig{figure= photo.eps,angle=0,clip=,width=9.5cm}}
\vspace{5mm} \centerline{\large this will be used for a photo}
\end{figure}

\end{document}